\documentclass[twocolumn,showpacs,preprintnumbers,amsmath,amssymb]{revtex4}

\usepackage{graphicx}% Include figure files
\usepackage{dcolumn}% Align table columns on decimal point
\usepackage{bm}% bold math

\newcommand{\blueflag}[1]{{\color{blue} #1}}
\usepackage[colorlinks=true,linktocpage=true,linkcolor=blue,citecolor=blue]{hyperref}

\begin{document}

%\preprint{}

\title{Functional renormalization group study \\of the Quark-Meson model with $\omega$ meson}% Force line breaks with \\

\author{Hui Zhang}
\email{Mr.zhanghui@mails.ccnu.edu.cn}
\author{Defu Hou}
\email{houdf@mail.ccnu.edu.cn}
\author{Toru Kojo}
\email{torujj@mail.ccnu.edu.cn}
\author{Bin Qin}
\email{qinbin@mails.ccnu.edu.cn}
 
\affiliation{Institute of Particle Physics (IOPP) and Key Laboratory of Quark and Lepton Physics (MOE),  Central China Normal University, Wuhan 430079, China}

\date{\today}

%%%%%%%%%%%%%%%%
\begin{abstract}

We study the phase diagram of two-flavor massless QCD at finite baryon density by applying the functional renormalization group (FRG) for a quark-meson model with $\sigma, \pi$, and $\omega$ mesons. The dynamical fluctuations of quarks, $\sigma$, and $\pi$ are included in the flow equations, while the amplitudes of $\omega$ fields are also allowed to fluctuate. At high temperature the effects of the $\omega$ field on the phase boundary are qualitatively similar to the mean-field calculations; the phase boundary is shifted to the higher chemical potential region. As the temperature is lowered, however, the transition line bends back to the lower chemical potential region, irrespective to the strength of the vector coupling. In our FRG calculations, the driving force of the low temperature first order line is the fluctuations rather than the quark density, and the effects of $\omega$ fields have little impact. At low temperature, the effective potential at small $\sigma$ field is very sensitive to the infrared cutoff scale, and this significantly affects our determination of the phase boundaries. The critical chemical potential at the tricritical point is affected by the $\omega$-field effects but its critical temperature stays around the similar value.  Some caveats are given in interpreting our model results.

\end{abstract}

\maketitle

%%%%%%%%%%Section1%%%%%%%%%%
\section{\label{sec:Intro} Introduction}

The phase diagram of Quantum Chromodynamics (QCD) has been of great interest to theoretical and experimental researches \cite{ref:Munzinger, ref:Friman}. While it became possible to study the high temperature region quantitatively due to experimental studies and the lattice Monte-Carlo simulations, our understanding for the phase diagram at high baryon density remains uncertain, partly because the lattice simulations are not directly applicable due to the infamous fermion-sign problem \cite{ref:deForcrand}, and also because the nuclear interactions at finite density are very complex. But in recent years a lot of hints to understand the phase structure have become available thanks to the experimental efforts such as Beam Energy Scan (BES) program at RHIC \cite{Luo:2017faz}, the constraints from the lattice QCD \cite{Bazavov:2017dus}, and astrophysics at very low temperature \cite{Baym:2017whm}.

A schematic quark model description at high baryon density, typically based on the Nambu-Jona-Lasinio or quark-meson models, has been also developed and several results beyond the mean field treatments are available \cite{ref:Schaefer1, ref:Schaefer2, ref:Herbst, ref:Strodthoff1, ref:Aoki, ref:Fukushima,Kamikado:2012cp}. One of the methods to go beyond the mean field (MF) is the functional renormalization group (FRG), which efficiently includes various fluctuation effects in the strongly correlated system. It is known that the fluctuation effects can change the order of the phase transitions, and thereby can be very important in understanding the QCD phase diagram.

Typically the FRG is applied to quark models of two-flavors with the scalar ($\sigma$) and pseudo scalar ($\pi$) fluctuations \cite{ref:Schaefer1, ref:Schaefer2, ref:Herbst}. There are also studies for the vector ($\rho$) and axial-vector ($a_1$) fluctuations in the isovector channels \cite{ref:Rennecke, ref:Eser, ref:Jung}. On the other hand, to the best of our knowledge, the $\omega$-fluctuations were taken into account only in the context of the Walecka type nucleon-$\sigma$-$\omega$ models whose main target is the nuclear matter at low temperature and density \cite{ref:Drews1, ref:Drews2, ref:Drews3}. In the quark model context, the mean field of the $\omega$ meson is known to have the significant impact on the phase boundary and the location of the critical end point \cite{ref:Fukushima2, Lourenco:2012yv, Bratovic:2012qs}, so it is natural to examine the stability of the mean-field picture against the $\omega$-fluctuations. In this paper we will take into account the $(\sigma,\pi,\omega)$-fluctuations and study their impacts on the phase diagram.

In this paper we focus on the phase diagram for the massless two-flavor QCD. Typically, in the chiral limit including fluctuations, the phase diagram is of the second order at high temperature and low chemical potential, and of the first order at low temperature and high chemical potential. There exists a tri-critical point where the second order line changes into the first order one \cite{ref:Schaefer2, ref:Lu, ref:Adhikari, ref:Herbst2}. We checked our calculations by reproducing this feature.

The structure of this paper is as follows. In Sec.~\ref{sec:model} we introduce our model and summarize the framework of the mean field approximation and the FRG method. In Sec.~\ref{sec:results} we examine the fluctuation effects and their impacts on the phase boundaries. Sec.~\ref{sec:summary} is devoted to summary.

%%%%%%%%%%Section2%%%%%%%%%%
\section{\label{sec:model} The Quark-Meson model with $\omega$ meson}

The Lagrangian of the two-flavors Quark-Meson model with $\omega$ meson in Minkowski space is
\begin{eqnarray}\label{eq:Lag}
		\mathcal L=&&\bar\psi\Big[i\gamma_\mu\partial^\mu-g_s(\sigma+i\gamma_5\,\boldsymbol\tau\cdot\boldsymbol\pi)-g_v\gamma_\mu\omega^\mu + \mu \gamma_0 \Big]\psi \nonumber\\
		&&+\tfrac 12\partial_\mu\sigma \partial^\mu\sigma+\tfrac 12\partial_\mu\boldsymbol\pi\cdot\partial^\mu\boldsymbol\pi -\tfrac 14 F_{\mu\nu}F^{\mu\nu} \nonumber\\
		&&- U(\sigma, \boldsymbol{ \pi}, \omega ),
\end{eqnarray}
with the field strength tensor $F_{\mu\nu} = \partial_\mu\omega_\nu - \partial_\nu\omega_\mu$. A field $\psi$ is the light two flavor quark field $\psi=(u, d)^T$. A bold symbol stands for a vector, and $\boldsymbol\tau=(\tau_1, \tau_2, \tau_3)$ are the Pauli matrices in isospin space. The potential for $\sigma$, $\pi$, and $\omega$ is
\begin{eqnarray}\label{eq:U}
U(\sigma, \boldsymbol{\pi},\omega ) 
=  \frac{\lambda}{4}(\sigma^2 + \boldsymbol{\pi}^2 -f_\pi^2 )^2  -\frac{\, m_v^2 \,}{2} \omega_\mu\omega^\mu \,,
\end{eqnarray}
where $f_\pi$ is the pion decay constant. We use the value $f_\pi = 93\,{\rm MeV}$, although its value in the chiral limit should be slightly smaller, $\simeq 87\, {\rm MeV}$.

The parameters in our model are $g_s, g_v, m_v$, and $\lambda$. The values of these parameters can differ for the MF and FRG calculations when we try to reproduce the same value for quantities such as the constituent quark mass of $\sim 300\,{\rm MeV}$. As for the value of $g_v$ and $m_v$, in our calculations they always appear in the form of $g_v/m_v$, so we will not discuss their values independently. Typical values in our problem are $m_v \sim 1\,{\rm GeV}$ and $g_v$ is about $\sim 1-10$, so the range of $g_v/m_v \simeq 10^{-3}$-$10^{-2}\, {\rm MeV}^{-1}$ is the natural choice in our model.
%With the explicit symmetry breaking, there are also parameters $\nu^2$ and $H$. But in this paper the explicit symmetry breaking case is discussed only for the MF analyses in which the forms of $\nu^2$ and $H$ are fixed by the low energy theorem; $\nu^2=f_\pi^2-m_\pi^2/\lambda$ and  $H=m_\pi^2 f_\pi$ with the pion decay constant $f_\pi$ and the pion mass $m_\pi$ which should be fixed to the empirical values.

%%%%%%%%%%%%
\subsection{\label{sec:MF} Mean-field approximation}

The chiral symmetry of the vacuum is explicitly broken and the expectation values of the meson fields are $\langle\sigma\rangle=f_\pi$ and $\langle\boldsymbol\pi\rangle=0$. Due to the rotational symmetry, only the zero-component of the vector field $\omega_\mu$ can have an expectation value~\cite{ref:Floerchinger}. Only considering the time component $\omega_0$ of the vector field $\omega_\mu$, the mean field potential reads as
\begin{eqnarray}\label{eq:U}
U_{ {\rm MF} } (\sigma,\omega_0 ) 
=   \frac{\lambda}{4}(\sigma^2  -f_\pi^2 )^2 - \frac{\, m_v^2 \,}{2} \omega_0^2 \,.
\end{eqnarray}
The mean-field effective potential is
\begin{equation}\label{eq:UT}
\Omega_{ {\rm MF} }= \Omega_{\bar{\psi}\psi } + U_{ {\rm MF} } (\sigma,\omega_0) \,,
\end{equation}
with the thermal quark and antiquark contributions ($\mu$: quark chemical potential; $T$: temperature; $\beta=1/T$)
\begin{eqnarray}
 \Omega_{\bar{\psi}\psi } = 
 &-&\nu_q \int\frac{\mathrm{d^3}\boldsymbol p}{(2\pi)^3} \bigg\{E_q \theta(\Lambda_{ {\rm MF} }^2-\boldsymbol p^2) \bigg\} \nonumber\\
 &-&\nu_q T\int\frac{\mathrm{d^3}\boldsymbol p}{(2\pi)^3}
\bigg\{\ln[1+e^{- \beta(E_q-\mu_{ {\rm eff} })}] \nonumber\\
&+& \ln[1+e^{- \beta (E_q+\mu_{ {\rm eff} }) }] \bigg\} ,
\end{eqnarray}
where $\nu_q$ is the degeneracy factor $\nu_q=$2(spin)$\times$ 2(flavor) $\times$ 3(color) $=12$ and $E_q=\sqrt{\boldsymbol p^2+m_{ {\rm eff} }^2}$. The first term is the fermion vacuum fluctuation contribution; if we dropped it off the transition in the chiral limit would be always the first order \cite{ref:Skokov}.
The effective quark (antiquark) mass and chemical potential are given as
\begin{equation}
m_{ {\rm eff} }=g_s \sigma,~~~~~\mu_{ {\rm eff} }=\mu-g_v \omega_0 .
\end{equation}
For a given $T$ and $\mu$, the gap equation for $\omega_0$ can be derived by %finding the extrema of the mean-field effective potential Eq.~(\ref{eq:UT}),
solving the quantum equation of motion for $\omega_0$,
\begin{equation}
\omega_0 = {g_v\over m_v^2} n(T, \mu-g_v \omega_0),
\end{equation}
which is the self-consistent equation. Here the quark density $n$ is determined by
\begin{equation}
n(T,\mu-g_v\omega_0)=- \,{\partial \over \partial\mu} \Omega_{\bar{\psi}\psi}(T,\mu-g_v\omega_0) \,.
\end{equation}
At this level, the vector coupling $g_v$ and the mass of the $\omega_0$ field are not independent; $g_v \omega_k$ is proportional to $(g_v/m_v)^2$.  Only their ratio $g_v/m_v$ appears in both MF and FRG calculations. 

In our calculation we follow the choice of Ref.~\cite{ref:Scavenius} and set the parameters $g_s = 3.3$ and $\lambda = 20$, with which the constituent quark mass in vacuum is $M_{\rm vac}=g_s f_\pi \simeq 307 \, {\rm MeV}$ and the sigma mass is $m_\sigma=\sqrt{2\lambda f_\pi^2 } \simeq 588\, {\rm MeV}$.

It should be remembered that in the MF calculations the strength of $\omega_0$ fields is proportional to quark number density $n$. Once we include the fluctuations, however, the quark number density is given by the sum of single particle contribution {\it plus} the contributions from other fluctuations, so such a proportionality relation no longer holds.

%%%%%%%%%%%%%%%%
\subsection{\label{sec:FRG} FRG flow equation}

The functional renormalization group (FRG) is a powerful non-perturbative tool in quantum field theories and statistical physics \cite{ref:Berges} and has been widely applied to QCD effective models \cite{ref:Schaefer1, ref:Schaefer2, ref:Tripolt, ref:Tripolt1, ref:Herbst, ref:Strodthoff2}. The effective average action $\Gamma_k$ with a scale $k$ obeys the exact functional flow equation %derived by C.~Wetterich and T.~R.~Morris in 1993.
\begin{equation}\label{eq:Wettericheq}
\partial_k \Gamma_{k}= \tfrac 12\operatorname{Tr}\bigg[\frac{\partial_k R_{k}}{\, \Gamma_{k}^{(2)}+R_{k} \,} \bigg] ,
\end{equation}
where $\Gamma_k^{(2)}$ is the second functional derivative of the effective average action with respect to the fields. The trace includes a momentum integration as well as traces over all inner indices. An infrared regulator $R_k$ is introduced to suppress fluctuations at momenta below the scale $k$.

In this study the dynamical fields in the flow equation are quarks, $\sigma$, and $\pi$, and they affect the effective potential and the size of $\omega_0$-fields. 
Unlike the spatial components of vector fields, the $\omega_0$ fields are not dynamical because it does not couple to the time derivative. Therefore the value of $\omega_0$ is completely fixed by specifying the values of other fields.
At each scale $k$ in the flow equation, we determine the value of $\omega_0$ fields by solving the consistency equation for given $\sigma$ and $\boldsymbol{\pi}$, so the resultant $\omega_0$ may be written as $\omega_{0,k}(\sigma, \boldsymbol{\pi} )$. This $\omega_{0,k}$ field in turn appears in the effective chemical potential for quarks, affecting the dynamical fluctuations in the flow equations. Throughout our study we neglect the flow of all wave-function renormalization factors.

The scale-dependent effective potential can be expressed by replacing the potential $U$ with the scale-dependent one $U_k$:
\begin{equation}\label{eq:Gamma_k}
\Gamma_k= \int \mathrm{d^4}x \;  \mathcal{L}|_{U\to U_k} ,
\end{equation}
with the Euclidean Lagrangian from Eq.~(\ref{eq:Lag}) for which the temperature is introduced by a Wick rotation to imaginary time $\int\mathrm d^4 x \equiv \int_0^{1/T}\mathrm d x_0 \int_V d^3 x$. Due to the chiral symmetry, the potential $U$ depends on $\sigma$ and $\pi$ only through the chiral invariant
\begin{equation}
\phi^2 \equiv \sigma^2 + \pi^2 \,.
\end{equation}
Starting with some ultraviolet (UV) potentials $U_\Lambda$ as our initial conditions, we integrate fluctuations and obtain the scale dependent $U_k$, which is artificially separated into the $\omega$-independent and dependent terms,
\begin{equation}\label{eq:U_k}
U_k=U_k^\phi +U_k^\omega \,,
\end{equation}
where the function form of $U_k^\phi$ will be determined without assuming any specific forms, while for the potential of the $\omega$-field we keep using the same form as in Eq.~(\ref{eq:Lag}),
\begin{equation}
U_k^\omega=-\tfrac 12 m_v^2 \omega_{0,k}^2 \, .
\end{equation}
Later we will also perturb our results by allowing $\omega^4$-terms, and check that our results are not significantly affected.

With these setup, we follow the standard methods to compute the FRG. For the computation of the flow equation, there are some freedom to choose the regulator $R_k$. We use the $3d$-analogue of the optimized regulator, which was proposed by Litim~\cite{ref:Litim},
\begin{eqnarray}
R_{k,B}(\boldsymbol p)&&=(k^2-\boldsymbol p^2)\theta(k^2-\boldsymbol p^2), \\
R_{k,F}(\boldsymbol p)&&=-\boldsymbol{p}\cdot\boldsymbol\gamma\left(\sqrt{\frac{k^2}{\boldsymbol p^2}} -1\right)\theta(k^2-\boldsymbol p^2) , \label{eq:regulators}
\end{eqnarray} 
for bosons and fermions respectively. Inserting Eq.~(\ref{eq:Gamma_k}-\ref{eq:regulators}) into Eq.~(\ref{eq:Wettericheq}), the flow equation for the potential $U_k^\phi$ can be obtained as
\begin{eqnarray}\label{eq:flow}
	\partial_k U_k^\phi (T,\mu)=\frac {k^4}{12\pi^2} \bigg\{\frac{3[1+2n_{\text B}(E_\pi)]}{E_\pi}+\frac {1+2n_{\text B}(E_\sigma)}{E_\sigma} \nonumber\\
		-\frac{2\nu_q\big[1-n_{\text F}(E_q,\mu^k_{\text{eff}})-n_{\text F}(E_q,-\mu^k_{\text{eff}})\big]}{E_q}\bigg\} ,
\end{eqnarray}
with single-particle energies are
\begin{eqnarray}
&& \label{expression1}
E_\pi=\sqrt{k^2 + 2 U_k' }   \,,\\
&& \label{expression2}
E_\sigma=\sqrt{k^2 + 2 U_k' + 4\phi^2 U_k'' } \,, \\
&& \label{expression3}
E_q=\sqrt{k^2+g_s^2\phi^2} \,,
\end{eqnarray}
 for a pion, sigma-meson, and quark, respectively; we also defined $U_k' \equiv \partial U_k/\partial \phi^2$. Here the mass terms are given as usual definition $m_\pi^2 = \delta^2 \Gamma/\delta \pi^2$, etc., while we found it convenient to use expressions (\ref{expression1}), (\ref{expression2}), (\ref{expression3}) in our equations. We have assigned the $\sigma$ quantum number in the radial direction for the effective potential, and the $\pi$ quantum number for the other directions. Note that during the FRG evolution pions may have the finite mass, as $U_k'$ can be nonzero for general $\phi$, vanishing only at the stationary point. 

The effective chemical potential, $\mu^k_{\text{eff}}=\mu-g_v\,\omega_{0,k}$, depends on the scale $k$ through $\omega_{0,k}$. The boson and fermion occupation numbers are
\begin{equation}
	n_B(E)=\frac {1}{\, e^{\beta E}-1 \,} \,,~~ n_F(E,\mu)=\frac {1}{\, e^{\beta(E-\mu)}+1 \,} \,.
\end{equation}

Apparently, the flow equation should be solved in the $\phi$ and $\omega_0$ directions. But fields $\omega_0$ are not dynamical, so the flow equation of $\omega_0$ fields can be computed for a given value of $\phi$, like the Gauss law constraint in gauge theories. At each momentum scale $k$, we determine $\omega_{0,k}$ by solving
\begin{equation}
	\frac{\, \partial U_k \,}{\, \partial\omega_{0,k} \,}=0 .
\end{equation}
The dependence on the $\omega_{0,k}$ manifestly appears only through the mass term and the fermion loop; we have a relation $m_v^2 \omega_0 \sim g_v \langle \bar{\psi} \gamma_0 \psi \rangle \sim \partial \Gamma_{ {\rm fermion} }/\partial \mu$. 
The RG evolution of this relation yields the flow equation
\begin{equation}\label{eq:omegak}
	\partial_k\,\omega_{0,k}=-\frac{2g_v\, k^4}{\, \pi^2m_v^2 E_q \,} \frac{\partial}{\, \partial\mu \,}\left( n_{\text F}(E_q,\mu^k_{\text{eff}})+n_{\text F}(E_q,-\mu^k_{\text{eff}}) \right) .
\end{equation}
This equation, together with Eq.(\ref{eq:flow}), constitutes our flow equations for the effective potential $U_k$ and the $\omega_k$-field as functions of $\phi$.

Note that the flow equation for $\omega_0$ can be solved for a given $\phi$, independently of the potential $U^\phi_k$ (which only tells us where the minimum of $\phi$ is). Thus, in our numerical calculations we first calculate $\omega_{0,k}$ as a function of $\phi$. Then the resultant $\omega_{0,k}(\phi)$ will be used in the FRG evolution equation (\ref{eq:flow}) for $U^\phi_k (\phi) $. 

To understand the behavior of $\omega_0$, for the moment we ignore the $k$-dependence in $\mu_{ {\rm eff} }$ in Eq.(\ref{eq:omegak}), and carry out the integration over $k$. Then the resulting expression for $\omega_0$ is some factor times the MF expression for the number density. But unlike the MF case, $\omega_0$ is not directly proportional to the physical number density, because the baryon density gets contributions not only from single particles but also fluctuations (see Eq.\ref{nb}). Moreover, as we will see in Sec.\ref{sec:orderpara}, if we include the $k$-dependence in $\mu_{ {\rm eff} }^k$, the $\omega^k_0$ field at $k_{ {\rm IR} }$ is not even proportional to the single particle contribution. Therefore the extrapolation of the MF relation $\omega_0 \sim n$ does not work at all to understand the FRG results.

Finally, the initial conditions for the flow equations must be set up. The UV scale $\Lambda$ should be sufficiently large in order to take into account the relevant fluctuation effects and small enough to render the description in terms of the model degrees of freedom realistic~\cite{ref:Drews1}.  In our calculation we follow the choice of the Ref.~\cite{ref:Schaefer2}, $\Lambda=500\, {\rm MeV}$. The initial condition for the potential is
\begin{equation}
	U^\phi_\Lambda =\tfrac\lambda 4 \phi^4 \,, %\rho^2 ,
	\label{eq:IC}
\end{equation}
and set the parameters $g_s=3.2$, $\lambda=8$ with the vacuum effective potential from the FRG computation having the minimum at $\sigma_{ {\rm vac} } \simeq 93\, {\rm MeV}$, which is regarded as $f_\pi$. 
We note that the value of $\lambda$, which enforces $\phi$ to stay near $f_\pi$, is considerably smaller than the MF case ($\lambda\sim 20$). If we start with another initial condition with an additional $\phi^2$ term to give the mass, we need to readjust $\lambda$ but obtain qualitatively similar results; in fact, starting with the condition Eq.(\ref{eq:IC}), the scale evolution first generates the $\phi^2$ terms, reflecting the universality.

The initial condition for the $\omega$ field has not been examined in detail, and we simply try
\begin{equation}
	\omega_{0,\Lambda}(\phi) = 0 \,.
\end{equation}
Later we will also present the result of another different initial condition, but it will turn out that such modification does not change the main story in this paper.

Assembling all these elements, we calculate the effective potential with the fluctuations integrated to $k_{ {\rm IR} } = 0$. The final step is to find $\phi =\sigma^*$ which minimizes the effective potential. At the minimum the effective potential is identified as the thermodynamic potential,
\begin{equation}
\Omega (\mu, T) = \Gamma_{ k_{ {\rm IR} } = 0 } (\mu, T, \sigma^*) \,.
\end{equation}
In practice, it is numerically expensive to reduce the IR cutoff, and we typically stop the integration around $k_{ {\rm IR} } \simeq 10\,{\rm MeV}$. The baryon number density is then obtained by taking the derivative with respect to $\mu_B = N_{ {\rm c} } \mu$,
\begin{equation}
n_B (\mu, T) = - \frac{1}{\, N_{ {\rm c} } \,} \frac{\partial \Gamma_{ k_{ {\rm IR} } = 0 } (\mu, T, \sigma^*) }{\partial \mu}\,.
\label{nb}
\end{equation}
The derivative is taken numerically with the interval $\Delta \mu = 0.1\, {\rm MeV}$.

%%%%%%%%%%Section3%%%%%%%%%%
\section{\label{sec:results} Results}
%%%%%%%%%%%%%%%%%%%%%%%%

%%%%%%%
%\begin{figure}[thb]
%\includegraphics[width=240pt]{MF_PD_without_Vac}
%\caption{\label{fig:MF_PD} The mean-field $T-\mu$ phase diagram \comment{without vacuum fluctuation ($\Lambda_{MF}=0$)} for different vector couplings \comment{in} the explicit symmetry breaking case. Solid lines show the first order phase transitions. Dots show the critical end point (CEP), star shows the vanishing of the CEP.}
%\end{figure}
%%%%%%%

%%%%%%%%%%%%%%%%%
\subsection{\label{sec:MFresults} The mean-field results}

%We briefly summarize the MF results. Fig.~\ref{fig:MF_PD} is the $T-\mu$ phase diagram of mean-field calculations \comment{without vacuum fluctuation ($\Lambda_{MF}=0$)} \comment{in} the explicit chiral symmetry breaking case for different vector couplings. There is a crossover phase transition at high temperature and small chemical potential, and a first order phase transition at low temperature and large chemical potential. Critical end points ($T_c,\mu_c$) are shown. As $g_v/m_v$ increase, the critical end point move to the right bottom side of the phase diagram. The critical end point vanishes at $g_v/m_v =5.3\times 10^{-3}\, {\rm MeV^{-1} }$.  This result is consistent with Ref.~\cite{ref:Fukushima2, Lourenco:2012yv, Bratovic:2012qs}.

%%%%%%%
%\begin{figure}[thb]
%\includegraphics[width=240pt]{MF_PD_ChiralLimit_without_Vac}
%\caption{\label{fig:MF_massless} The mean-field $T-\mu$ phase diagram \comment{without vacuum fluctuation ($\Lambda_{MF}=0$)} for the two-flavor massless QCD with different vector couplings. The phase boundary is the first order transition. }
%\end{figure}
%%%%%%%

%For the chiral limit case \comment{without vacuum fluctuation ($\Lambda_{MF}=0$)} (Fig.~\ref{fig:MF_massless}), however, the phase transition is always of the first order. The transition chemical potential gets larger with increasing vector couplings for a fixed temperature. 

%%%%%%%
\begin{figure}[thb]
\includegraphics[width=240pt]{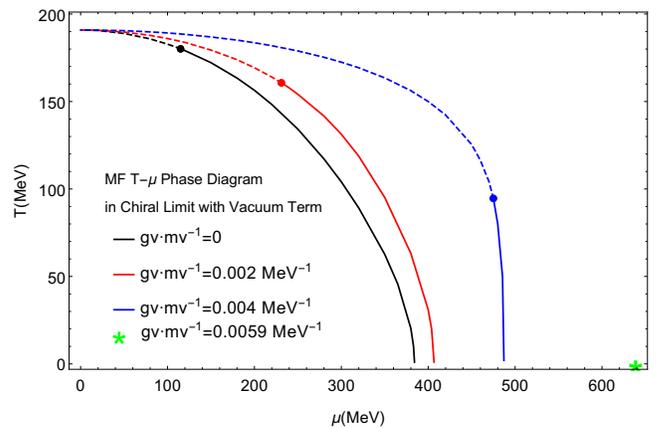}

\caption{\label{fig:MF_Vac} The mean-field $T-\mu$ phase diagram including vacuum fluctuation ($\Lambda_{ {\rm MF} }=260\, {\rm MeV}$) for the two-flavor massless QCD with different vector couplings. Solid lines show the first order phase transitions, dashed lines show the second order phase transition. Dots show the tricritical point (TCP), star shows the vanishing of the TCP.}
\end{figure}
%%%%%%%

We briefly summarize the MF results for the chiral limit. Following Ref.~\cite{ref:Skokov}, we include the fermion vacuum term with $\Lambda_{ {\rm MF} }=260\, {\rm MeV}$. Without this term the phase boundary is always the first order. With the vacuum term, there is a second order phase transition at high temperature and small chemical potential, and a first order phase transition at low temperature and large chemical potential. At tricritical points (TCP) with ($T_c,\mu_c$) the order of the phase transition changes. As $g_v/m_v$ increase, the TCP moves to the right bottom side of the phase diagram, and eventually vanishes at $g_v/m_v =5.9\times 10^{-3}\, {\rm MeV^{-1} }$.

%\comment{In Ref.~\cite{ref:Skokov}, the authors show that ``the fermion vacuum fluctuations can change the order of the phase transition in the chiral limit and strongly influence physical observables". We also show this feature. And the fermion vacuum fluctuation can make the transition temperature lager for low chemical potential. For the chiral limit case with fermion vacuum contribution (Fig.~\ref{fig:MF_Vac}), there is a second order phase transition at high temperature and small chemical potential, and a first order phase transition at low temperature and large chemical potential. Tricritical points ($T_c,\mu_c$) are shown. As $g_v/m_v$ increase, the tricritical point move to the right bottom side of the phase diagram. The tricritical point vanishes at $g_v/m_v =5.9\times 10^{-3}\, {\rm MeV^{-1} }$.}

%%%%%%%
\subsection{\label{sec:FRGresults} The results of the FRG}

\begin{figure}[thb]
\includegraphics[width=240pt]{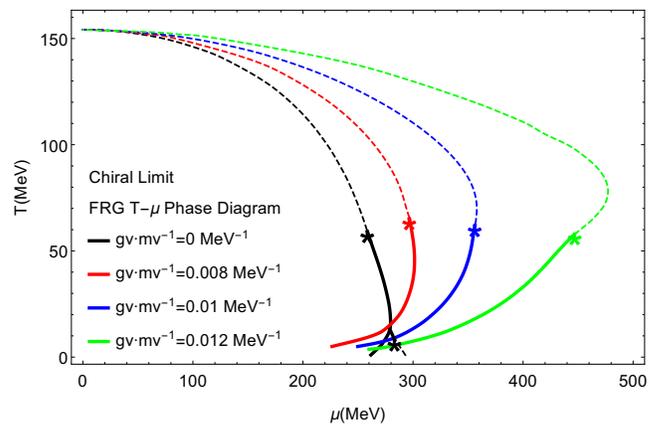}
\caption{\label{fig:FRG_FD} The phase diagram of the FRG with different vector couplings. Dashed (solid) lines show the second (first) order phase transition. Stars show the tri-critical end point (TCP).}
\end{figure}

In this section we present the FRG results for the chiral limit. The phase diagrams for different coupling constants are summarized in Fig.~\ref{fig:FRG_FD}.
Here we give a quick summary of the results before dictating the details of calculations: 
%(i) At high temperature the fluctuation effects \comment{(unlike MF calculation, FRG includes not only the fermion vacuum fluctuation but also the meson vacuum fluctuation and higher order thermal fluctuation)} turn the first order line in the MF calculations (for the chiral limit \comment{without fermion vacuum contribution} case, see Fig.~\ref{fig:MF_PD}) into the second order, yielding the TCP for all vector couplings which we have examined; (ii) 
(i) While the critical chemical potential of the TCP is sensitive to the vector coupling, its critical temperature is similar for different vector couplings; 
(ii) At high temperature, the vector couplings shifts the phase boundaries to higher chemical potential as in the MF, but the curves strongly bend back toward lower temperatures irrespective to the value of $g_v$; the curves with different vector couplings approach one another. We note that the back bending behavior has already been found in other FRG calculations without the vector coupling \cite{ref:Schaefer2, ref:Strodthoff1, ref:Herbst, ref:Aoki1, ref:Weyrich, ref:Aoki, ref:Tripolt1}. 
%\comment{Especially in Ref.~\cite{ref:Tripolt1}, this behavior is discussed in detail.}

Behavior (ii) is somewhat unexpected to us: what we initially expected was that the vector coupling tempers not only the growth of the number density but also fluctuations, so the results should be similar to the MF results which do not have the back bending behavior. Our FRG calculations, however, do not follow this expectation; as we will examine later, the fluctuation effects develop even before the appearance of the quark Fermi sea, affecting the phase structure before the vector coupling becomes important.

We have checked that the result at $g_v=0$, in which case the phase boundary has another TCP at low temperature and high chemical potential, is consistent with Ref.\cite{ref:Schaefer2, ref:Herbst2}. Below we shall examine more details---such as the behavior of effective potentials, order parameters, baryon density---to understand the structure of our phase diagram at finite vector couplings. 

%%%%%%%%%%%
\subsubsection{\label{sec:Effpot} The effective potentials}
%%%%%%%%%%%

\begin{figure}[thb]
\includegraphics[width=240pt]{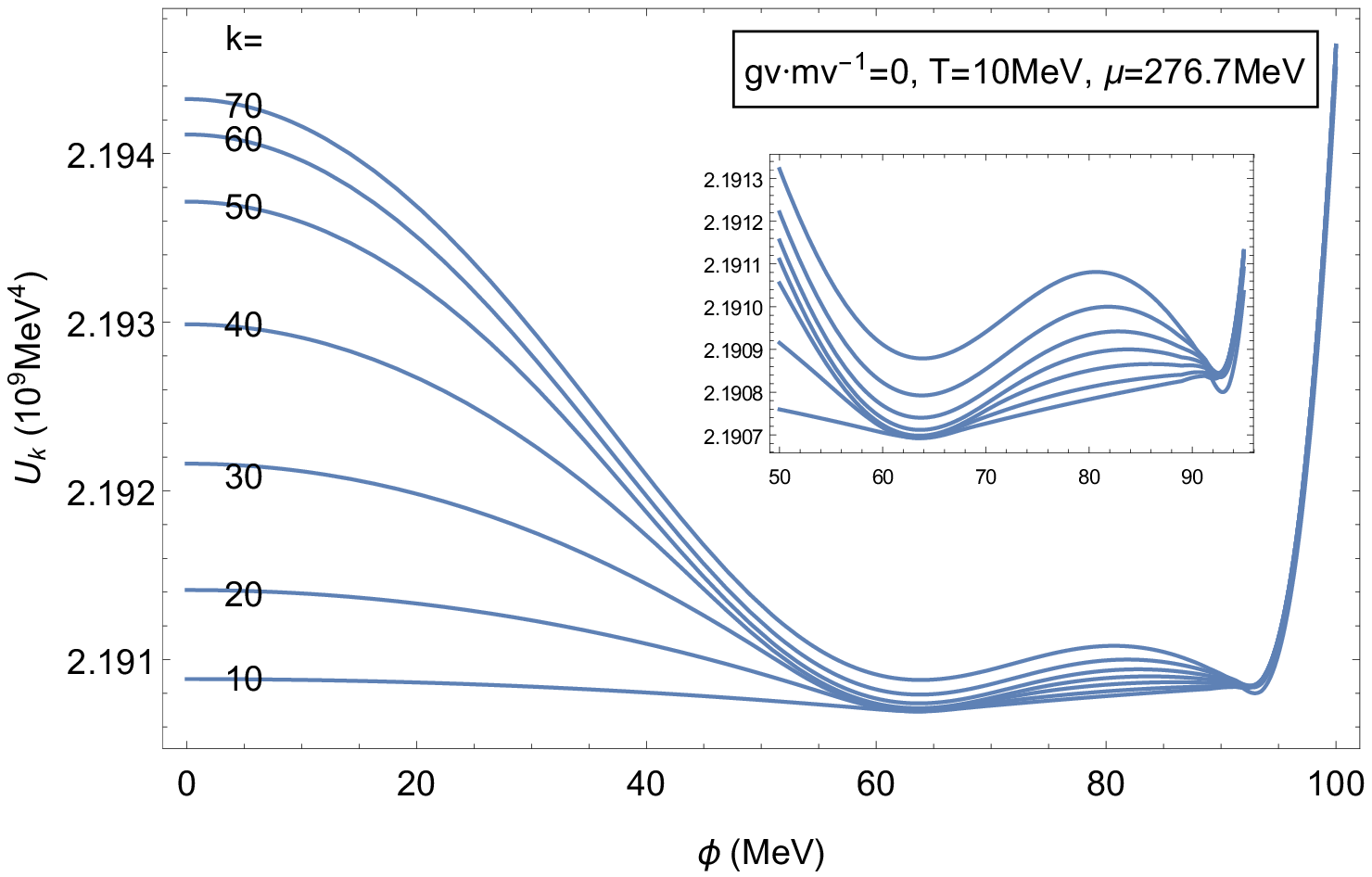}
\includegraphics[width=240pt]{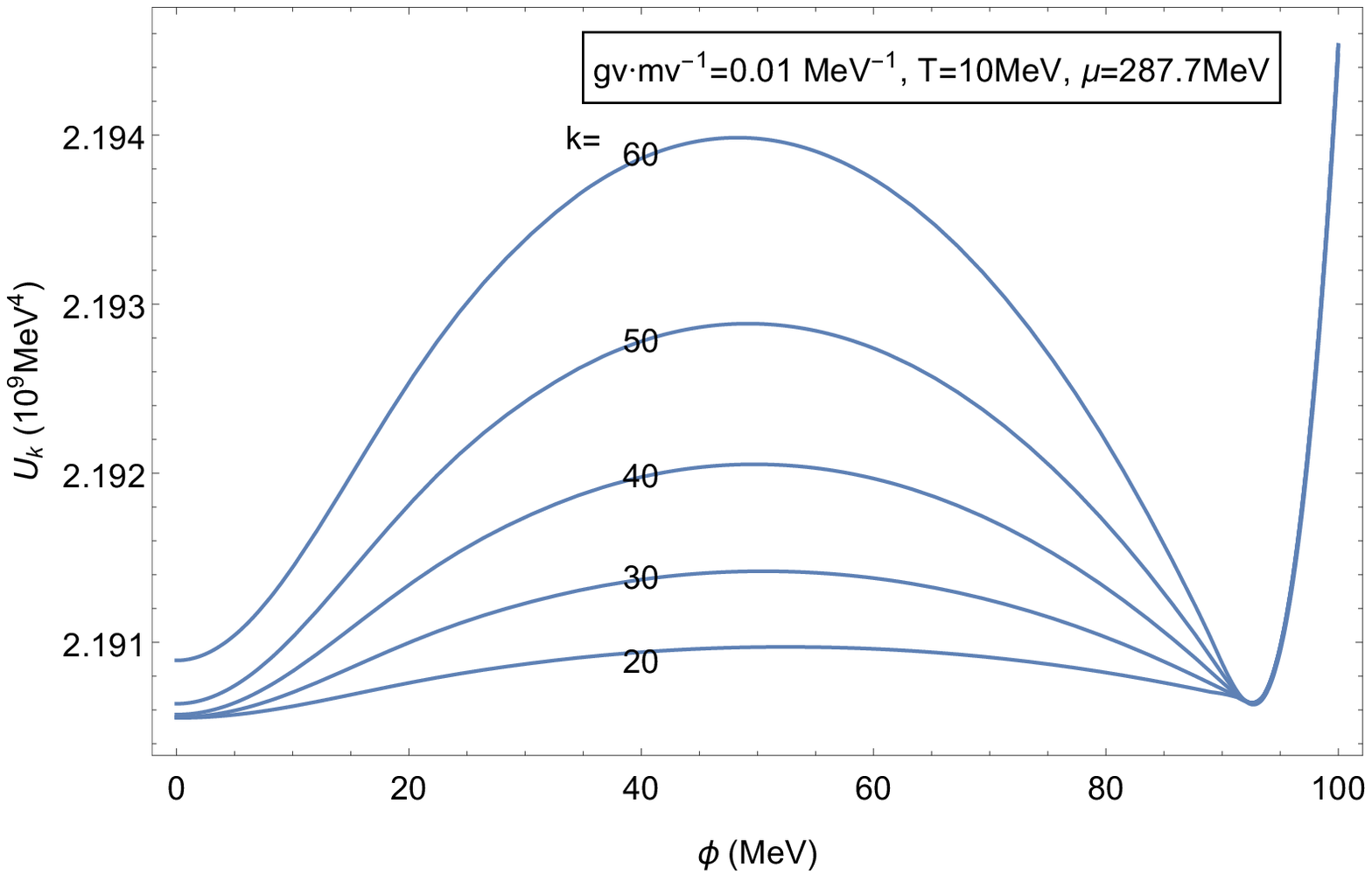}
\caption{\label{fig:U-phi} 
The scale evolution of the effective potential $\Gamma_k(\phi)$ at low temperature.  We compare the results with and without the vector coupling near the phase boundaries in the FRG results; (top) $g_v/m_v=0$, $T=10\, {\rm MeV}$ and $\mu=276.7\, {\rm MeV}$; (bottom) $g_v/m_v=0.01\, {\rm MeV}^{-1}$, $T=10\, {\rm MeV}$ and $\mu=287.7\, {\rm MeV}$.  }
\end{figure}

The full flow equation~(\ref{eq:flow}) is solved on a grid \cite{ref:Adams, ref:Schaefer1}. To check the stability of our numerical results, we compare two different methods to solve the flow equation. We got the same results for the fourth order Backward Differentiation Formula (BDF) and the Linearly Implicit Midpoint method. The flow equation is integrated from the UV momentum $k=\Lambda = 500\,{\rm MeV}$ to the IR momentum $k_{ {\rm IR} }=10-20\, {\rm MeV}$ until the location of the minimum of the effective potential is stabilized (See Fig.~\ref{fig:U-phi}). The fluctuations erase the barrier between two local minima in the mean field potential, making the effective potential convex, as they should. 

Fig.~\ref{fig:U-phi} illustrates the evolution of the effective potential $\Gamma_k(\phi)$ towards the IR for different vector couplings. We fix the temperature to $T=10\,{\rm MeV}$ and choose the chemical potential near the phase boundaries of the FRG results. The top panel is the result for $g_v=0$ at $ \mu=276.7\, {\rm MeV}$, and the bottom one is for $g_v/m_v=0.01\, {\rm MeV}^{-1}$ at $\mu=287.7\, {\rm MeV}$.

We first examine the case without the vector coupling. Before integrating the fluctuations out, the global minimum stays around $\phi \simeq f_\pi$ as in the vacuum case. With fluctuations, while they hardly affect the effective potential near $\phi \simeq f_\pi$, they crucially affect the effective potential at lower $\phi$. Below $k\simeq 70\, {\rm MeV}$, the local minimum around $\phi =60-70\,{\rm MeV}$ becomes the global one. Therefore the fluctuations let the phase transition occur at lower chemical potential than the MF case: the local minimum at $\phi = 60-70\,{\rm MeV}$ turns into the global minimum at $\mu \simeq 276.7\, {\rm MeV}$, long before the local minimum at $\phi \simeq 93\,{\rm MeV}$ merges into the global minimum at $\phi =60-70\,{\rm MeV}$. Therefore the phase transition is of the first order in this case. As we increase $\mu$, the global minimum smoothly approaches the local minimum at $\phi =0$, leading to the second order phase transition at $\mu\simeq 282\,{\rm MeV}$ (see Fig.~\ref{fig:FRG_FD}). This also means that there exists a tri-critical end point. All these features are consistent with the calculations in Ref.\cite{ref:Schaefer2}. 

At finite vector coupling, many features remain similar as the $g_v=0$ case (except the appearance of global minima in the $g_v=0$ case). In short, the fluctuations do not modify the effective potential around a local minimum at $\phi \simeq 93\,{\rm MeV}$, while the potential around $\phi \simeq 0$ is reduced significantly by fluctuations. This feature is common for all vector couplings in our study. In the next section we will examine this feature in more detail.

It is important to notice that the minimum around $\phi \simeq 0$ is very sensitive to the IR cutoff scale $k_{ {\rm IR} }$, as one can see from Fig. \ref{fig:U-phi}. This means that at $\phi\simeq 0$ there are strong fluctuations with small excitation energies. If we had stopped integrating the fluctuations before the results are stabilized, the minimum at $\phi \simeq f_\pi$ would remain the absolute minimum, resulting in very different phase boundaries which are closer to the MF results. 

From the second derivative of the FRG effective potential, we can obtain the $\sigma$ mass which depends on the scale and the order parameter. We evaluate the vacuum value of the $\sigma$ mass at the global minimum $\sigma_{\rm vac}\simeq 93 {\rm MeV}$ of the potential in the IR, and find it is about 303 MeV with the parameters $g_s=3.2, \lambda=8$.

%%%%%%%%%%%
\subsubsection{\label{sec:orderpara} Order parameter and baryon density}
%%%%%%%%%%%

%
\begin{figure}[htb]
\includegraphics[width=240pt]{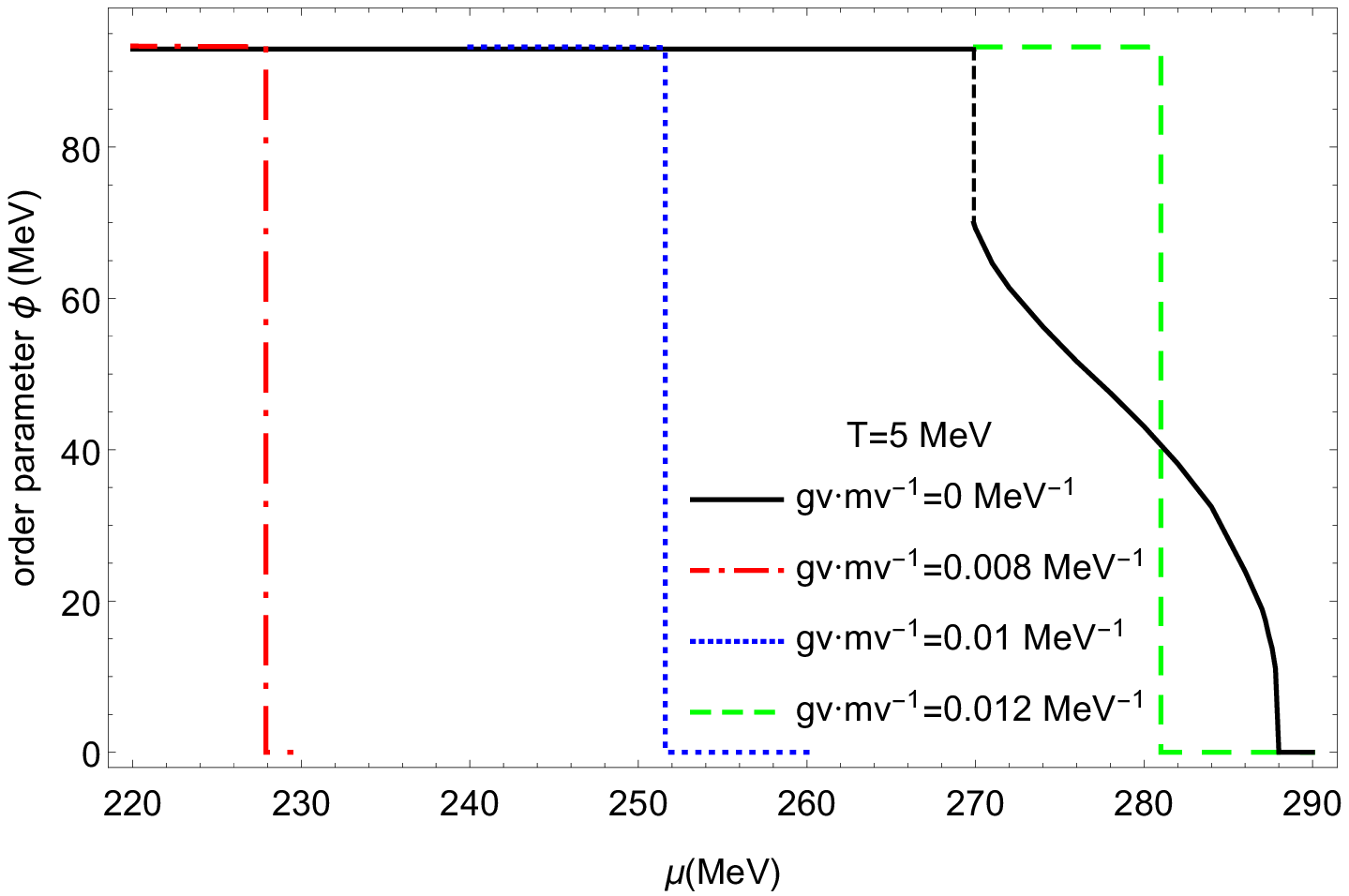}
\includegraphics[width=240pt]{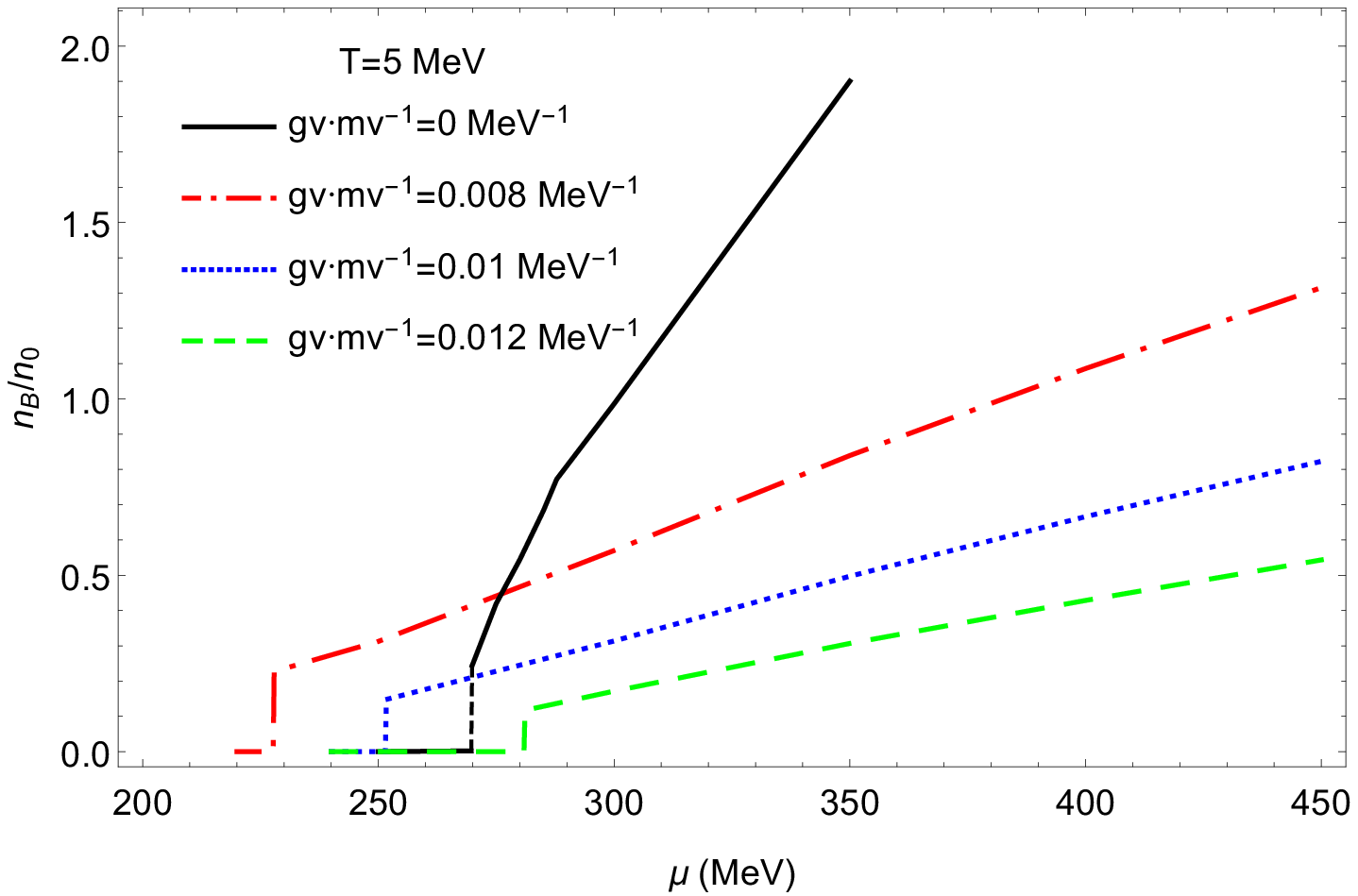}
\caption{\label{fig:phi-mu_T5} The vacuum expectation value of the order parameter $\phi$ and the baryon density as a function of chemical potential $\mu$ at $T=5\, {\rm MeV}$, calculated by the FRG with different vector couplings. The $g_v=0$ case has the first order transition around $\mu \simeq 270\,{\rm MeV}$, and then the second order phase transition around $\mu \simeq 290\,{\rm MeV}$. The other cases have only the first order phase transitions.}
\end{figure}

To examine the phase structure in more detail, we check the behavior of the order parameter $\phi$ and the baryon density, especially their relationship. 

We first examine the results at $T=5\,{\rm MeV}$, Fig.~\ref{fig:phi-mu_T5} for (top) the order parameter and (bottom) the baryon density normalized by the nuclear saturation density $n_0 = 0.16\,{\rm fm}^{-3}$. %As we already noted,
The result of the $g_v =0$ case has the first order phase transition at $\mu \simeq 270\,{\rm MeV}$ and the second order phase transition at $\mu \simeq 288\,{\rm MeV}$. The other cases $g_v/ m_v = (0.8, 1.0, 1.2)\times 10^{-2} \, {\rm MeV}^{-1}$ all have the first order phase transitions. After the transition the vector coupling tempers the growth of the baryon density, as we originally expected.

It seems that the change in order parameter is not driven by the baryon density. This is in contrast to typical MF calculations in which the baryon density develops first, and then drives the reduction of the chiral order parameter. Thus, the mechanism of the chiral restoration found in our calculations for $T\simeq 5\,{\rm MeV}$ is very different from the conventional density driven one; in fact the phase transition occurs before $\mu$ reaches the vacuum effective quark mass ($g_s f_{\pi} \simeq 298\,{\rm MeV}$).

One might think that the jumps in baryon density in the FRG calculations are conceptually similar to what was suggested in the self-bound quark matter hypothesis; the quark matter is more stable than the nuclear matter so that the quark matter can appear before the baryon chemical potential reaches the nucleon mass \cite{Witten:1984rs,Bodmer:1971we}. But in our calculations the baryon density just after the emergence of matter is at most $n_B \sim n_0$, presumably too low for the quark matter picture to be justified.

\begin{figure}[htb]
\includegraphics[width=240pt]{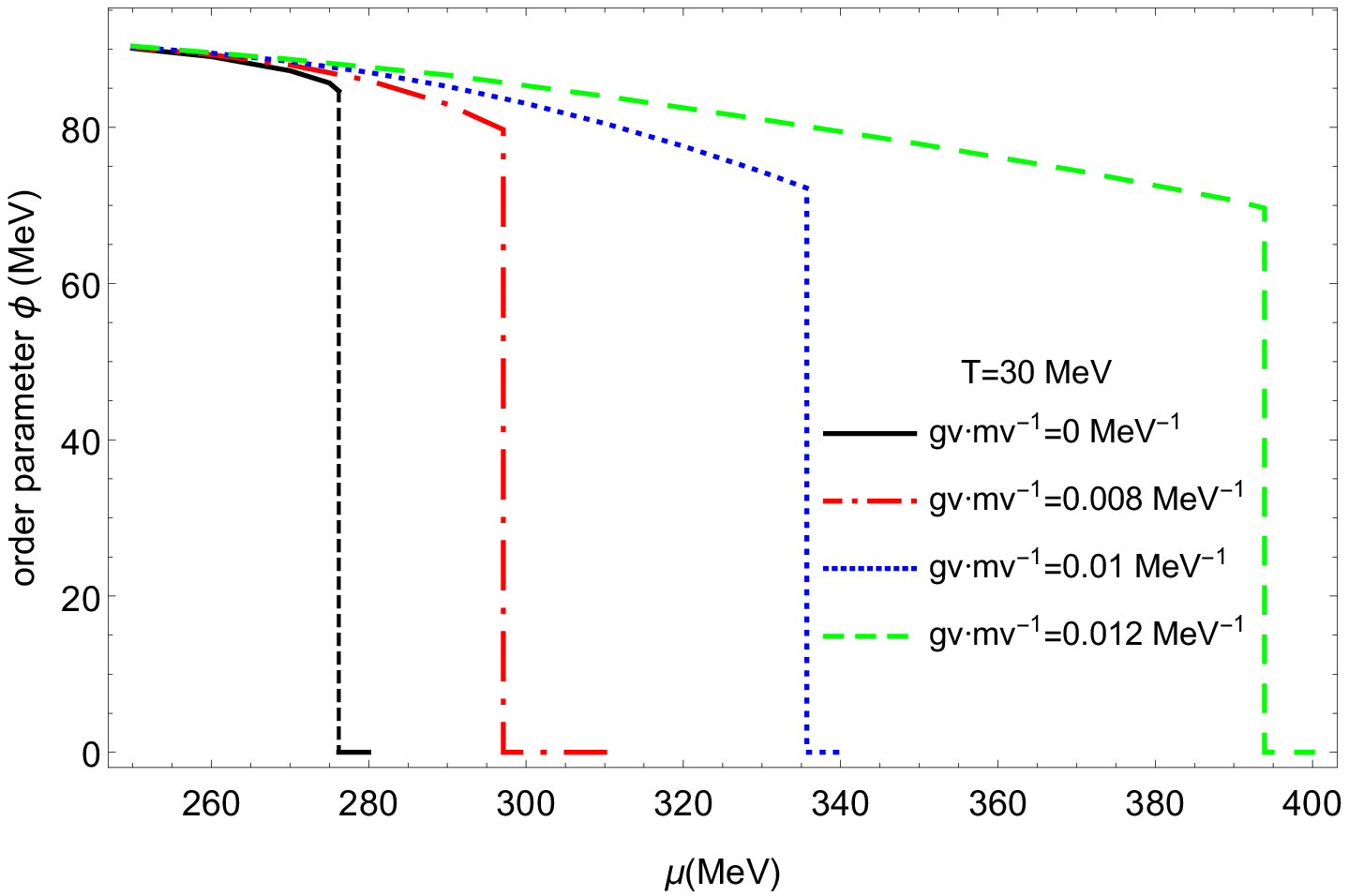}
\includegraphics[width=240pt]{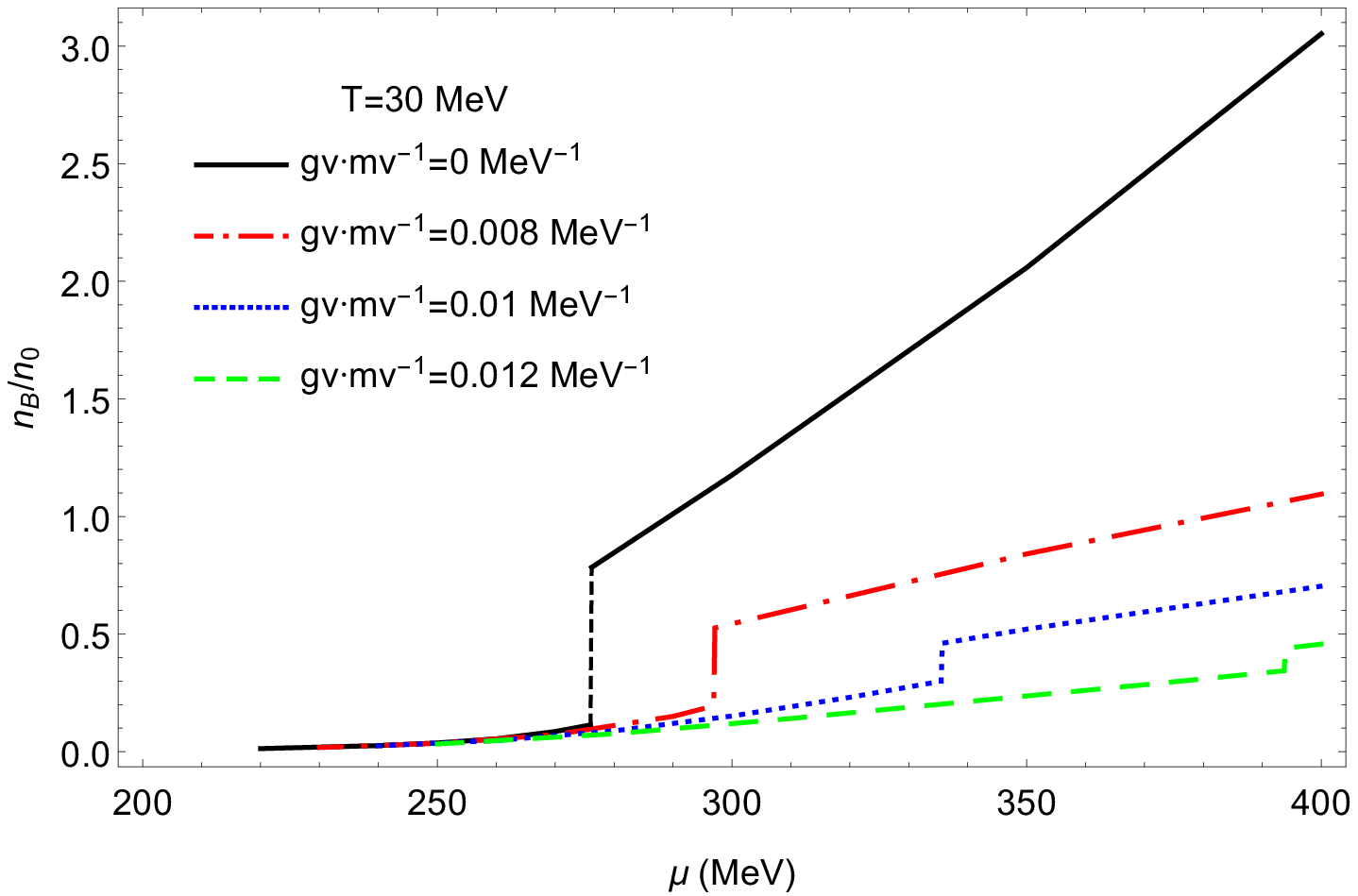}
\caption{\label{fig:phi-mu_T30} The same as Fig.~\ref{fig:phi-mu_T5}, except the temperature is now $T=30\,{\rm MeV}$.}
\end{figure}

Next we examine the results at $T=30\,{\rm MeV}$ in Fig.~\ref{fig:phi-mu_T30}. Compared to the $T=5\,{\rm MeV}$ case, the result is much closer to the MF behavior; the baryon density gradually develops and then the chiral restoration occurs. But still there remains the back bending behavior in the phase boundaries for all the vector couplings.

The $\mu$-dependence of the baryon density considerably deviates from $\sim \mu^3$ behavior expected from the single particle contributions. In fact, our derivation of the baryon density includes not only the fermionic but also the bosonic fluctuations which also depend on $\mu$, and somewhat unexpectedly the latter is more important especially when the vector coupling is large.

\begin{figure}[htb]
\includegraphics[width=240pt]{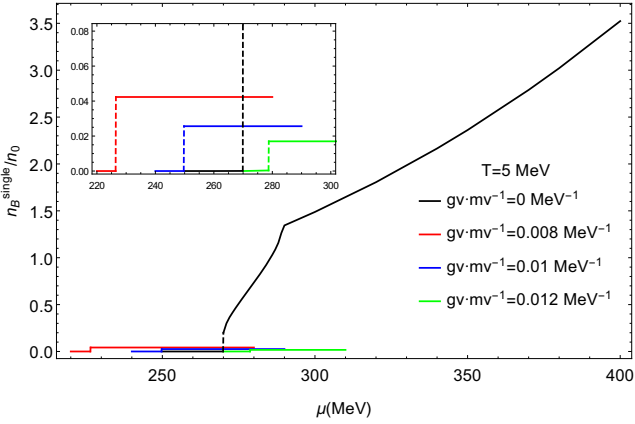}
\caption{\label{fig:nBF-mu} 
The baryon density of fermion part $n_{B}^{ {\rm single} }$ as a function of chemical potential $\mu$ for $T=5\, {\rm MeV}$ from the FRG with different vector couplings.}
\end{figure}

For further inspections, the baryon density from single particle contribution for $T=5\,{\rm MeV}$ is plotted in Fig.~\ref{fig:nBF-mu}. 

Without the vector coupling constant, baryon density from the fermion part $n_{B}^{ {\rm single} }$ approaches the $\sim \mu^3$ behavior. Actually the single particle contribution $n_{B}^{ {\rm single} }$,
\begin{eqnarray}
n_{B}^{ {\rm single} }
&&\equiv \frac{\, \nu_q T \,}{3} 
\int\frac{\mathrm{d^3}\boldsymbol p}{(2\pi)^3} \bigg\{\frac{1}{\, e^{(E_q^k-\mu^k_{ {\rm eff} })/T} +1\,} \nonumber\\
&&- \frac{1}{\, e^{(E^k_q+\mu^k_{ {\rm eff} })/T} + 1 \,} \bigg\}   \bigg|_{k=k_{ {\rm IR} } } \,,
\end{eqnarray}
 is larger than the total baryon density $n_B=n_{B}^{ {\rm single} } + n_{B}^{ {\rm fluct} }$ for large chemical potential, meaning that $n_{B}^{ {\rm fluct} } < 0$. 

\begin{figure}[htb]
\includegraphics[width=240pt]{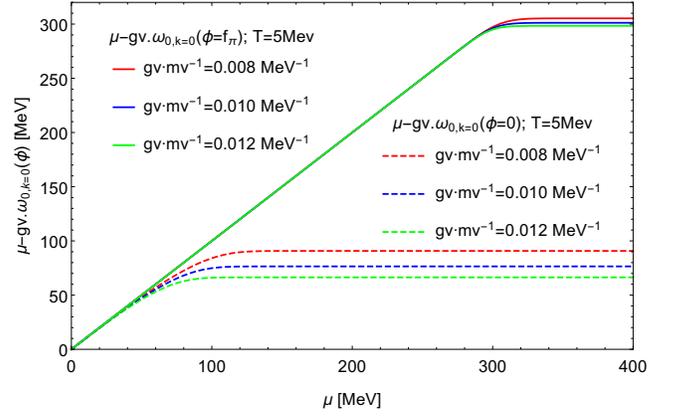}
\caption{\label{fig:mu_eff} Effective chemical potential $\mu-g_v\cdot \omega_{0,k=0}(\phi=0)$ as a function of chemical potential at fixed $T=5\ {\rm MeV}$ with different vector couplings.}
\end{figure}
\begin{figure}[htb]
\includegraphics[width=240pt]{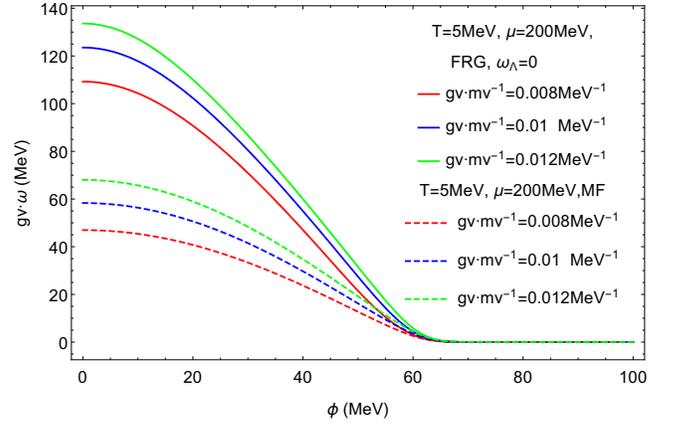}
\caption{\label{fig:omega-phi} Solutions $g_v\cdot \omega$ as a function of the chiral condensate $\phi$ at fixed $\mu=200\ {\rm MeV}, T=5\ {\rm MeV}$ with different vector couplings. Solid lines are for FRG results, and dashed lines for MF results.}
\end{figure}

In contrast, with non-vanishing vector coupling constants, the single particle contribution is significantly suppressed and the baryon density is almost saturated by fluctuation contributions after the first order phase transition happens (see Figs.~\ref{fig:nBF-mu} and~\ref{fig:mu_eff}), so that the single particle contribution remains small. Note that $\omega_0$ is large in spite of small baryon density; the MF-like relation $\omega_0 \propto n$ does not work at all. This means that the large amplitude of $\omega_0$ is induced by fluctuations rather than the quark density, as in the first order phase transition (see Fig.~\ref{fig:omega-phi}). While $\omega_0$ is large, the amplitudes of $g_v \omega_0$ does not exceed $\mu$ so that $\mu_{ {\rm eff} }$ does not reach a negative value.

\begin{figure}[htb]
\includegraphics[width=240pt]{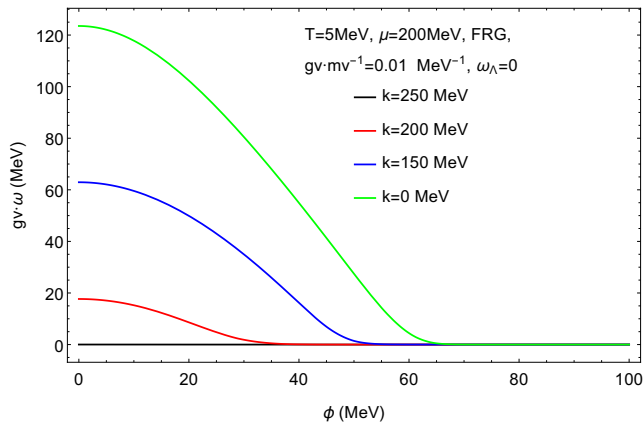}
\caption{\label{fig:omega-phi_k} The scale evolution of the omega field $g_v\cdot \omega$ as a function of the chiral condensate $\phi$ at fixed $\mu=200\ {\rm MeV}, T=5\ {\rm MeV}$ with fixed vector coupling constant $g_v/m_v=0.01\ {\rm MeV^{-1}}$.}
\end{figure}

We also plot the scale evolution of the omega field $g_v\cdot \omega$ as a function of the chiral condensate $\phi$ at fixed $\mu=200\ {\rm MeV}, T=5\ {\rm MeV}$ with fixed vector coupling constant $g_v/m_v=0.01\ {\rm MeV^{-1}}$ in Fig.~\ref{fig:omega-phi_k}. One can easily find that for small $\phi$ the omega field grows faster and faster as the scale decreases, but for large $\phi$ it stays zero.

\begin{figure}[htb]
\includegraphics[width=240pt]{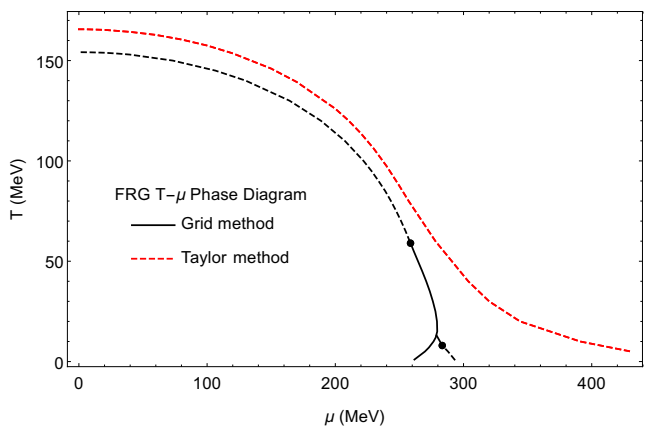}
\caption{\label{fig:FRG_PD_LPA} A comparison of the phase diagrams calculated by the grid method FRG and the Taylor methods. Dashed lines show the second order phase transition. The vector coupling is omitted for simplicity.}
\end{figure}
%

%%%%%%%%%%%%%%%%%
\subsection{\label{sec:others} Several other checks}

To check the stability of  our results, in this section we perturb our setup for calculations and try to identify the universal features.

%%%%%%%%%%%%%%%%%
\subsubsection{\label{sec:LPA} Truncated potential for $\phi$ }

Our FRG results in the previous section are very sensitive to the fluctuations. Here we focus on the effect of the $\phi$-fluctuations by using the Taylor expansion of $U_k$,
\begin{equation}\label{eq:U_LPA}
	U_k(\phi)= \frac{\, \lambda_k \,}{4} (\phi^2-a_{k})^2,
\end{equation}
with the scale dependent parameters $a_k$ and $\lambda_k$. We recall that our previous calculations did not assume any functional form for $U_k(\phi)$. Thus the difference from the Taylor method clarifies the importance of higher order vertices for $\phi$. We also emphasize that in this method there is, by construction, only one minimum at given $T$ and $\mu$, whose location is determined by the scale evolution of $a_k$. As before, the initial condition at $k_{{\rm UV}} = \Lambda$ is chosen to be $a_{ k =\Lambda} = 0$, $\lambda_{ k =\Lambda} = 15.2$ to reproduce the vacuum pion decay constant of $f_\pi \simeq 93\, {\rm MeV}$.

Within this simple approximation, we get the Taylor method $T-\mu$ phase diagram. Here we omitted the vector coupling. We found that the phase transition line is the second order everywhere and there is no back bending behavior (See Fig.~\ref{fig:FRG_PD_LPA}).

%%%%%%%
\subsubsection{\label{sec:discussion_on} The fourth order vector coupling \\constant and initial condition for $\omega$}

\begin{figure}[htb]
\includegraphics[width=240pt]{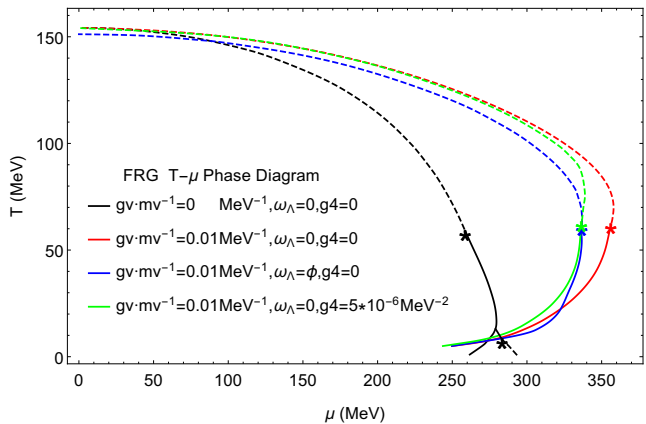}
\caption{\label{fig:FRG_FD_g4} Functional renormalization group $T-\mu$ phase diagram with different vector couplings and initial conditions for the omega meson.}
\end{figure}

We check the robustness of our results by varying treatments of the $\omega$-fields. From now on, $g_v/m_v$ is fixed to $0.01\, {\rm MeV}^{-1}$. We change the initial condition for the omega meson from $\omega_{k=\Lambda} =0$ to
\begin{equation}\label{eq:omega_Lambda}
    \omega_\Lambda=\phi \,.
\end{equation}
Starting with this initial condition, the value of $\omega$ as $k \rightarrow 0$ tends to a take larger value than the case with the initial condition $\omega_{\Lambda}=0$.
We found that this change tends to increase the value of $\omega$ at relatively large $\phi$, bringing the energy cost due to the repulsive force. As a result, the phase transition to $\phi=0$ occurs at lower temperature and chemical potential. But the overall structure of the phase diagram does not change, as seen in Fig.~\ref{fig:FRG_FD_g4}.

Next we also consider the effect of quartic coupling. Such repulsive quartic self-coupling is often introduced in the relativistic MF approach. We choose the form of the $\omega$ potential as
\begin{equation}
    U_k(\omega)=-\tfrac 12 m_v^2 \omega_{0,k}^2 +\tfrac{1}{12} g_4\cdot (g_v^2m_v^2)\cdot \omega_{0,k}^4 \,.
\end{equation}
With this configuration, the flow equations for $U_k$ and $\omega_{0,k}$ are both affected. We give the flow equation for $\omega_{0,k}$, which reads
\begin{eqnarray}\label{eq:omegak}
	\partial_k\,\omega_{0,k}=&&-\frac{2g_v\, k^4}{\, \pi^2m_v^2 E_q \,} \frac{1}{1-g_4 \cdot g_v^2 \omega_{0,k}^2} \nonumber\\
	&&\times \frac{\partial}{\, \partial\mu \,}\left( n_{\text F}(E_q,\mu^k_{\text{eff}})+n_{\text F}(E_q,-\mu^k_{\text{eff}}) \right) .
\end{eqnarray}
For the repulsive quartic term, we found it convenient to factor out $(g_v m_v)^2$ in writing the flow equation. Then $g_4$ has the mass dimension $-2$, and its natural size is $ \sim (1000\,{\rm MeV} )^{-2} \simeq 10^{-6} \,{\rm MeV}^{-2}$. 

We show the result for $g_4 = 5\times 10^{-6}\,{\rm MeV}^{-2}$ in Fig.~\ref{fig:FRG_FD_g4}. With the quartic term, the overall structure, such as the back bending behavior, is not significantly affected. The phase transition line shifts slight to the lower chemical potential region.

To summarize, the details of how we treat the $\omega$ meson part do not change the qualitative feature of the phase boundaries, at least for the natural range of model parameters.

%%%%%%%%%%Section4%%%%%%%%%%
\section{\label{sec:summary} Summary}

In this paper we discuss the quark meson model with $\sigma$, $\pi$, and $\omega$ mesons at finite temperature and density using the FRG. We focus on the effects of the $\omega$-mesons, which are known to be very important in MF determination of the phase boundaries.

Without $\omega$-fields, it has been known that FRG calculations typically lead to the back bending behavior at low temperature phase boundary. This behavior looks somewhat unnatural to us, and we expected that introduction of the repulsive density-density interactions would tame this problem. Our FRG results do not follow our expectation; what we found is that 
the low temperature first order phase transition in the FRG is induced by fluctuations, rather than number density as in the MF case, so that the structure of the low temperature boundaries remains similar for different values of vector couplings.

Another important finding in this study is that the effective potential at small $\phi$ is very sensitive to the infrared cutoff scale $k$. If we artificially stopped the integration before stabilizing the result, we would get very different phase boundaries. On the other hand, the results without going very small $k$ are closer to the conventional MF results which are easier to interpret on physical grounds. It is not clear to us whether there exist good rationales to ignore fluctuations in the very infrared.

We think that our FRG results show very strong fluctuation effects with which the results are hard to interpret. We believe that the problem of strong fluctuations should be solved in general context, without using specific features of QCD. Our model does not possess confinement, but the main sources in our fluctuations are color-singlet; so even after the successful modeling of confinement the issues of fluctuations are likely to remain. Further studies are called for.

A part of the origin of strong fluctuations may be our use of the chiral limit. It is known that even small current quark mass significantly increases the pion mass. Since our results on phase boundaries are very sensitive to the infrared scale $k$, the details of low-lying excitations should be important. Hence the obvious extension of the present study is to examine the impact of the explicit breaking. This should be discussed elsewhere.

 \blueflag{
 }

\begin{acknowledgments}
The authors thank J.~Liao and M.~Horvath for their helpful discussions. D.~H. and T.~K. acknowledge J.~Wambach for his comments on the back bending behavior during the workshop ``CPOD2017" held at the StonyBrook University.
We are grateful to K.~Redlich pointing out a fermion vacuum term overlooked in the mean-field calculations. We are indebted to J.~Pawlowski and D.~Rischke for valuable suggestions and discussions.
The work is supported in part by the Ministry of Science and Technology of China (MSTC) under the ``973" Project No. 2015CB856904(4) (D.~H.), and by NSFC under Grant Nos. 11375070, 11735007, 11521064 (D.~H.); 11650110435 (T.~K.).
H.~Z. gratefully acknowledges financial support from China Scholarship Council Grant No. 201706770051.
\end{acknowledgments}

%%%%%%%%%%Reference%%%%%%%%%%

\end{document}